\documentclass{article}

 \usepackage[final]{neurips_2021}
\usepackage{graphicx}
\setcitestyle{numbers,square}

\usepackage[utf8]{inputenc} 
\usepackage[T1]{fontenc}    
\usepackage{hyperref}       
\usepackage{url}            
\usepackage{booktabs}       
\usepackage{amsfonts}       
\usepackage{nicefrac}       
\usepackage{microtype}      
\usepackage{xcolor}         

\title{Beware of the Ostrich Policy: End-Users' Perceptions Towards Data Transparency and Control}

\author{%
  Sruthi Viswanathan \\
 Naver Labs Europe\\
  Meylan\\
  France \\
  \texttt{sruthi.viswanathan@naverlabs.com} \\
}

\begin{document}

\maketitle

\begin{abstract}
  End users' awareness about the data they share, the purpose of sharing that data, and their control over it, is key to establishing trust and eradicating privacy concerns. We experimented on personal data management by prototyping a Point-of-Interest recommender system in which data collected on the user can be viewed, edited, deleted, and shared via elements in the User Interface. Based on our qualitative findings, in this paper we discuss \textit{"The ostrich policy"} adopted by end users who do not want to manage their personal data. We sound a waking whistle to design and model for personal data management by understanding end users' perceptions towards data transparency and control. 
\end{abstract}

\section{Introduction}

The French expression \textit{``La politique de l'autruche''} roughly translates into \textit{“The ostrich policy”} in English. It describes the hypothetical behaviour of an ostrich to bury its head in sand, to avoid facing its fears. Despite the recent wave of awareness about data privacy and security, regulations such as General Data Protection Regulation (GDPR)~\cite{gdpr}, and documentaries such as The Great Hack~\cite{documentry}, our findings suggest that end users are practising the ostrich policy, by refusing to look at the available controls and/or the amount of data they are sharing with an everyday AI system. 

Today, the most profitable method to monetise web and mobile applications is via in-app advertisements. These advertisements rely on modelling users' personal preferences from the data collected about their online history. Location-based services providing information on Point of Interests (POIs) or transportation use a variety of personal data such as users' current location, preferences, routines, historical activities, etc., to provide successful recommendations. Users are found to exhibit different "disclosure styles" with differences in the kind and degree of personal data they are willing to disclose in return for services~\cite{bart}. On evaluating the design of Laksa, an intelligible mobile context aware application, users were found to want a bird’s eye view of the explanations on a first glance and then have access to more details on demand~\cite{laksa}.
 
\begin{figure}
  \centering
  \includegraphics[width=0.3\textwidth]{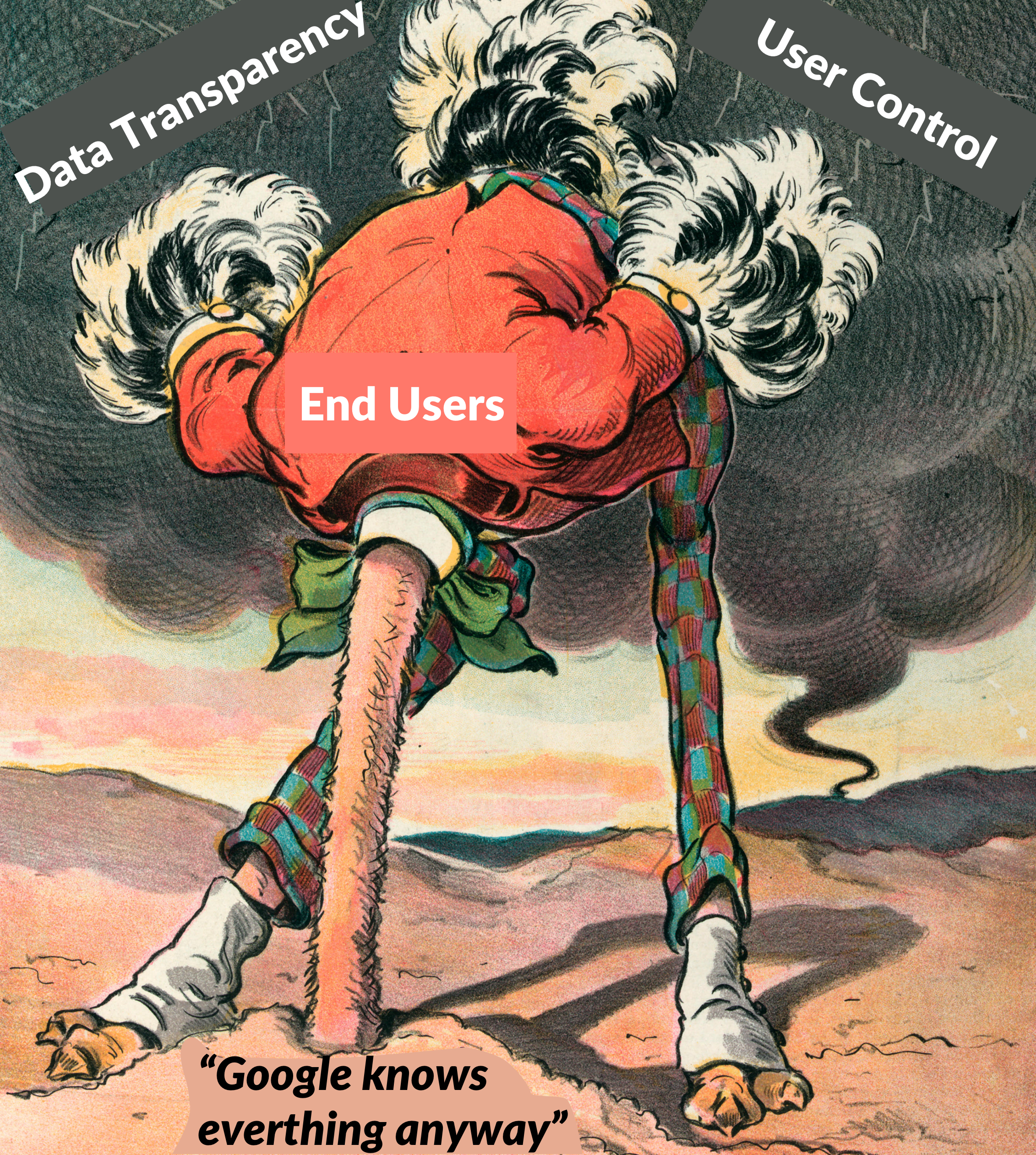}
  \caption{An illustration of end users adopting the ostrich policy towards personal data management}
   \label{fig:ostrich}
\end{figure}

Beyond understanding users' willingness to disclose and studying the kind, presentation, or interpretability of personal data, we sought to understand whether end-users' perceive data transparency and control to be useful and usable elements of an everyday AI system. In this short abstract, we report our qualitative observations from experimenting with a Wizard-of-Oz prototype of a POI recommender and discuss our surprise in finding end users to adopt the ostrich policy towards personal data management (as illustrated in Figure~\ref{fig:ostrich}).

 
 \newpage

\section{Experiment and Findings}
\begin{figure}
  \centering
  \includegraphics[width=0.8\textwidth]{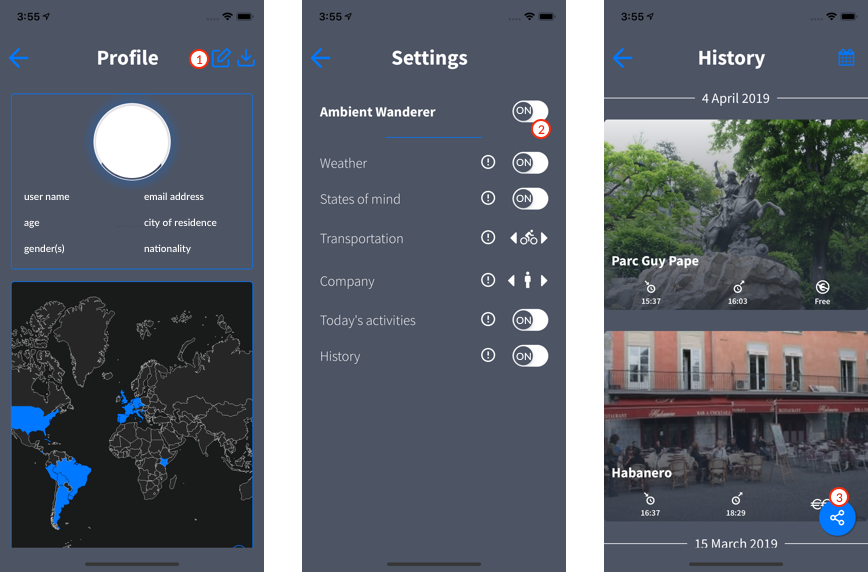}
  \caption{Screenshots of our prototype illustrating some of the simple controls that werer made available in the UI of our system to edit, delete, download (marked as 1), turn ON/OFF (marked as 2), or share (marked as 3) personal data}
   \label{fig:prototype}
\end{figure}

For our research project on designing Ambient Wanderer~\cite{aw} a personalised and contextualised POI recommender system, we prototyped a mobile application with interfaces illustrating the data shared by the user with our system (see Figure~\ref{fig:prototype}). Our settings tabs offered an overview of the contextual variables which inform the decisions made by the recommender algorithm. Users could also view detailed explanations on demand. With this transparent interface, we ran a hybrid Wizard-of-Oz experiment~\cite{hwoz} with 12 participants. Our participants were all young adults (age range: 19-40) who had recently relocated to Grenoble (a city in French Alps) and who were in the habit of frequently searching for POIs. With a pre questionnaire, we collected their personal data such as demographics and the details of POIs they had visited in Grenoble in the past few weeks. We displayed this data in the interface of our prototype during the experiment.  A log of the participant's past activities was shown under `History' and real-time information obtained via phone's sensors such as location, weather, and mode of transportation were shown under `Settings' (see Figure~\ref{fig:prototype}). We also used the personal data collected from the user and the contextual data at the time of their experiment to manually pick and recommend different POIs for every participant in a Wizard-of-Oz manner (i.e., no recommender engine was implemented in our prototype). For the scope of this paper, we only focus on the part of our study investigating the management of personal data. We met the participants in a café in the city centre and they were handed an iPhone with our prototype. Following think-aloud protocol~\cite{jaaskelainen2010think}, they interacted with transparent data by viewing, editing, granting or rejecting permissions, and discussed their views post interaction. A week after the experiment, we sent an online post-questionnaire to re-confirm their opinions and collect overall feedback about their experience. Data was collected and stored from this experiment was handled in compliance with the GDPR agreement~\cite{gdpr}. We audio recorded the sessions, transcribed the recordings, and thematically analysed~\cite{braun2012thematic} the transcripts. An agreement was reached on the resulting themes after five rounds of discussion between seven researchers (including the author of this paper) who were involved in the POI recommender project in our AI lab. All participants were rewarded with gift vouchers worth 30 Euros. In the rest of this section, we discuss our findings on end users' perceptions towards data transparency and control.

\subsection{Controlled Sharing: I choose when and what to share}
As we had expected, some of the participants (4/12) said that they would control the access to their data as per their need and comfort. A system that was "always ON" and constantly collecting data was perceived to be "stalking" by these participants. Therefore, they liked the choice of deciding when to allow/deny access and selecting what type data to share as in their comments below.

\begin{verbatim}
“I don’t like to be stalked and anytime I don’t need it I can turn off.”(P11)
\end{verbatim}

\begin{verbatim}
“I like it because I will choose(to turn ON and OFF).”(P10)
\end{verbatim}

Participants planned to creatively allow access if they want to record their data for their own need (such as for the purpose bookmarking) or deny access because they want to remove the effect of personalisation and seek serendipity as in their statements below.

\begin{verbatim}
“If I don’t want to record History I can turn it OFF. When I give permission 
it’s good for me because I can remember and find back (the previously 
visited POIs). If I don’t want I can remove it.”(P6)
\end{verbatim}

\begin{verbatim}
“If you want something new, I would turn this History OFF and look for 
options again (in the home screen).”(P1)
\end{verbatim}

Our observations suggest that some users find data transparency to be useful and are willing to use the choices available in the system for controling their personal data.

\subsection{Share Everything: Google knows everything anyway}

To our surprise, we found that most of the participants (8/12) sustained a – "What is the point?" attitude towards personal data management. To these participants, not thinking twice about the data they share was an easier choice to make as in their comments below.

\begin{verbatim}
“We are already tracked. Google knows where you go. It’s fine, Google knows
everything anyway, a little more is fine. I can also remember the places 
I visited (due to tracking history).”(P9)
\end{verbatim}

Some of the participants (4/8) who did not want any control over their data still found the transparent interface displaying their data logged by the system to be helpful as in their statements below.

\begin{verbatim}
“I don't want control but I don't mind seeing it, because it will be helpful 
if I like a place and want to go again. It is good for sharing an 
activity(POI) with a friend.”(P3)
\end{verbatim}

\begin{verbatim}
“I think this list can help. Some times you go to a place and months after that
you can't remember the name of that place. I won't select ON or OFF though.”(P7)
\end{verbatim}

While the rest of other participants (4/8) who did not want any control over their data, neither wanted any transparency, as commented by P5 below. 

\begin{verbatim}
“Maybe some people are afraid to share information with the app so they 
want to see what are sharing, I could be like– `meh'. I don't think not
having visibility is a problem for me.”(P5)
\end{verbatim}

We questioned them further and observed that this behaviour was connected to their anxiety about the large amount of data which they have been sharing so far with their everyday intelligent applications. P4 was willing to share their history of POIs visited with the system, however, they did not want to see what they share as stated below.

\begin{verbatim}
"It's just uncomfortable to look at. It is spooky, so I just want to hide 
it. I'm OK to share everything but don't show me what I'm sharing."(P4)
\end{verbatim}

Our findings suggest that most participants (8/12) considered it to be pointless to try and have control over the data they share with an everyday AI application and some participants (4/12) were uncomfortable to look at the amount of data they allowing access to.

\subsection{The aftermath: I had a change of heart}
\begin{figure}[ht]
  \centering
  \includegraphics[width=0.6\textwidth]{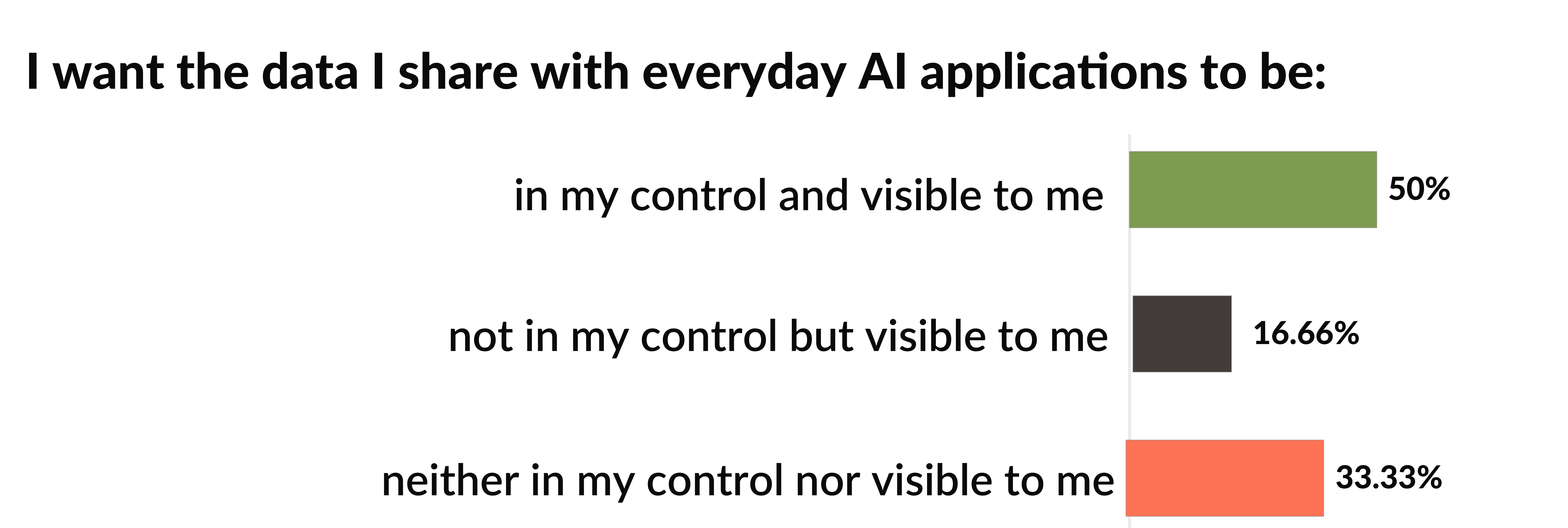}
  \caption{Post-questionnaire responses of our 12 participants about their perceptions towards user control and data transparency showing a change in the opinions of a few participants who moved from wanting only visibility of their data to wanting control as well}
   \label{fig:postQ}
\end{figure}

As the last part of our study, we followed up with the participants on the impact of our experiment with an online post questionnaire. As seen in Figure~\ref{fig:postQ}, half of the participants (6/12) wanted to have both control and transparency. This was different from our direct observation during which only 4/12 participants wanted both control and transparency. We found that two participants who wanted no control (only transparency) during the experiment, had changed their perceptions. Both these participants left open-ended comments in our post questionnaire explaining that they had a `changed of heart' about managing their personal data after participating in our experiment and discussing it with their friends and family.

\section{Discussion: The ostrich policy}
We observed that only half of our participants behaved as predicted in the literature~\cite{louisebark}, by liking the simple options provided in our system to handle their data. The choices of the other half of our participants who followed \textit{ The ostrich policy} by not wanting to look at their data and/or the available ways control it is understandable, as current everyday AI services (such as maps and video streaming) provide free and unlimited access in exchange for personal data. Although people are becoming reluctant to share their personal data and big-tech firms predict to abandon personalisation by 2025 due to the scarcity of personal data~\cite{marketers}, a serious design consideration is required to appeal to the group of end users who are likely to follow the ostrich policy. 

As harmful as end users' adoption of the ostrich policy is, we find it to be an interesting interdisciplinary problem to address. As seen in the initial insights from our post questionnaire, users following the ostrich policy might start to care more about their personal data based on the availability of controls and its ease of use. Designing data models, interfaces, and interactions with a dedicated effort to overcome the ostrich policy will pave the way for better data literacy, decisions, and control. Thus fostering a fair and ethical relationship between humans and AI for personal data management.

To conclude, despite our observations being limited to our sample and system, we believe that sharing our findings on \textit{ The ostrich policy} with researchers and practitioners working on Human-Centred AI (HCAI) will lead to interesting discussions and directions for Comprehensible AI (CAI) and Explainable AI (XAI).



\section{Acknowledgements}
I thank the team members of the Ambient Wanderer project for their continued discussions on \textit{ The ostrich policy} and all our participants for sharing their perceptions and providing us with their honest feedback. In particular, I wish to extend my special thanks to \textit{Jean-Michel Renders} for naming the above observed end-user's perceptions towards data management as \textit{“La politique de l’autruche”}.

\section{Disclaimer}
The opinions expressed in this paper are those of the author and do not necessarily reflect the opinions of the author's employer(s)

\bibliographystyle{ACM-Reference-Format}
\bibliography{ref}


\begin{thebibliography}{10}


\ifx \showCODEN    \undefined \def \showCODEN     #1{\unskip}     \fi
\ifx \showDOI      \undefined \def \showDOI       #1{#1}\fi
\ifx \showISBNx    \undefined \def \showISBNx     #1{\unskip}     \fi
\ifx \showISBNxiii \undefined \def \showISBNxiii  #1{\unskip}     \fi
\ifx \showISSN     \undefined \def \showISSN      #1{\unskip}     \fi
\ifx \showLCCN     \undefined \def \showLCCN      #1{\unskip}     \fi
\ifx \shownote     \undefined \def \shownote      #1{#1}          \fi
\ifx \showarticletitle \undefined \def \showarticletitle #1{#1}   \fi
\ifx \showURL      \undefined \def \showURL       {\relax}        \fi
\providecommand\bibfield[2]{#2}
\providecommand\bibinfo[2]{#2}
\providecommand\natexlab[1]{#1}
\providecommand\showeprint[2][]{arXiv:#2}

\bibitem[\protect\citeauthoryear{Amer and Jehane}{Amer and Jehane}{2019}]%
        {documentry}
\bibfield{author}{\bibinfo{person}{Karim Amer} {and} \bibinfo{person}{Noujaim
  Jehane}.} \bibinfo{year}{2019}\natexlab{}.
\newblock \bibinfo{title}{The Great Hack. United States: Netflix}.
\newblock
\newblock
\urldef\tempurl%
\url{https://www.imdb.com/title/tt4736550/}
\showURL{%
Retrieved August 17, 2021 from \tempurl}


\bibitem[\protect\citeauthoryear{Barkhuus and Dey}{Barkhuus and Dey}{2003}]%
        {louisebark}
\bibfield{author}{\bibinfo{person}{Louise Barkhuus} {and}
  \bibinfo{person}{Anind Dey}.} \bibinfo{year}{2003}\natexlab{}.
\newblock \showarticletitle{Location-Based Services for Mobile Telephony: a
  Study of Users' Privacy Concerns.}, Vol.~\bibinfo{volume}{2003}.
\newblock


\bibitem[\protect\citeauthoryear{Blum and Omale}{Blum and Omale}{2019}]%
        {marketers}
\bibfield{author}{\bibinfo{person}{Kelly Blum} {and} \bibinfo{person}{Gloria
  Omale}.} \bibinfo{year}{2019}\natexlab{}.
\newblock \bibinfo{title}{Gartner Predicts 80\% of Marketers Will Abandon
  Personalization Efforts by 2025}.
\newblock
\newblock
\urldef\tempurl%
\url{https://www.gartner.com/en/newsroom/press-releases/2019-12-02-gartner-predicts-80--of-marketers-will-abandon-person}
\showURL{%
Retrieved August 17, 2021 from \tempurl}


\bibitem[\protect\citeauthoryear{Braun and Clarke}{Braun and Clarke}{2012}]%
        {braun2012thematic}
\bibfield{author}{\bibinfo{person}{Virginia Braun} {and}
  \bibinfo{person}{Victoria Clarke}.} \bibinfo{year}{2012}\natexlab{}.
\newblock \showarticletitle{Thematic analysis.}
\newblock  (\bibinfo{year}{2012}).
\newblock


\bibitem[\protect\citeauthoryear{Commission}{Commission}{[n.d.]}]%
        {gdpr}
\bibfield{author}{\bibinfo{person}{European Commission}.}
  \bibinfo{year}{[n.d.]}\natexlab{}.
\newblock \bibinfo{booktitle}{\emph{2018 reform of EU data protection rules}}.
\newblock
\urldef\tempurl%
\url{https://ec.europa.eu/commission/sites/beta-political/files/data-protection-factsheet-changes_en.pdf}
\showURL{%
\tempurl}


\bibitem[\protect\citeauthoryear{J{\"a}{\"a}skel{\"a}inen}{J{\"a}{\"a}skel{\"a}inen}{2010}]%
        {jaaskelainen2010think}
\bibfield{author}{\bibinfo{person}{Riitta J{\"a}{\"a}skel{\"a}inen}.}
  \bibinfo{year}{2010}\natexlab{}.
\newblock \showarticletitle{Think-aloud protocol}.
\newblock \bibinfo{journal}{\emph{Handbook of translation studies}}
  \bibinfo{volume}{1} (\bibinfo{year}{2010}), \bibinfo{pages}{371--374}.
\newblock


\bibitem[\protect\citeauthoryear{Knijnenburg, Kobsa, and Jin}{Knijnenburg
  et~al\mbox{.}}{2013}]%
        {bart}
\bibfield{author}{\bibinfo{person}{Bart~P. Knijnenburg},
  \bibinfo{person}{Alfred Kobsa}, {and} \bibinfo{person}{Hongxia Jin}.}
  \bibinfo{year}{2013}\natexlab{}.
\newblock \showarticletitle{Dimensionality of information disclosure behavior}.
\newblock \bibinfo{journal}{\emph{International Journal of Human-Computer
  Studies}} \bibinfo{volume}{71}, \bibinfo{number}{12} (\bibinfo{year}{2013}),
  \bibinfo{pages}{1144--1162}.
\newblock
\showISSN{1071-5819}
\urldef\tempurl%
\url{https://doi.org/10.1016/j.ijhcs.2013.06.003}
\showDOI{\tempurl}


\bibitem[\protect\citeauthoryear{Lim and Dey}{Lim and Dey}{2011}]%
        {laksa}
\bibfield{author}{\bibinfo{person}{Brian~Y. Lim} {and}
  \bibinfo{person}{Anind~K. Dey}.} \bibinfo{year}{2011}\natexlab{}.
\newblock \showarticletitle{Design of an Intelligible Mobile Context-Aware
  Application}. In \bibinfo{booktitle}{\emph{Proceedings of the 13th
  International Conference on Human Computer Interaction with Mobile Devices
  and Services}} (Stockholm, Sweden) \emph{(\bibinfo{series}{MobileHCI '11})}.
  \bibinfo{publisher}{Association for Computing Machinery},
  \bibinfo{address}{New York, NY, USA}, \bibinfo{pages}{157–166}.
\newblock
\showISBNx{9781450305419}
\urldef\tempurl%
\url{https://doi.org/10.1145/2037373.2037399}
\showDOI{\tempurl}


\bibitem[\protect\citeauthoryear{Viswanathan, Omidvar-Tehrani, Bruyat,
  Roulland, and Grasso}{Viswanathan et~al\mbox{.}}{2020a}]%
        {aw}
\bibfield{author}{\bibinfo{person}{Sruthi Viswanathan},
  \bibinfo{person}{Behrooz Omidvar-Tehrani}, \bibinfo{person}{Adrien Bruyat},
  \bibinfo{person}{Fr\'{e}d\'{e}ric Roulland}, {and}
  \bibinfo{person}{Antonietta~Maria Grasso}.} \bibinfo{year}{2020}\natexlab{a}.
\newblock \showarticletitle{Designing Ambient Wanderer: Mobile Recommendations
  for Urban Exploration}. In \bibinfo{booktitle}{\emph{Proceedings of the 2020
  ACM Designing Interactive Systems Conference}} (Eindhoven, Netherlands)
  \emph{(\bibinfo{series}{DIS '20})}. \bibinfo{publisher}{Association for
  Computing Machinery}, \bibinfo{address}{New York, NY, USA},
  \bibinfo{pages}{1405–1418}.
\newblock
\showISBNx{9781450369749}
\urldef\tempurl%
\url{https://doi.org/10.1145/3357236.3395518}
\showDOI{\tempurl}


\bibitem[\protect\citeauthoryear{Viswanathan, Omidvar-Tehrani, Bruyat,
  Roulland, and Grasso}{Viswanathan et~al\mbox{.}}{2020b}]%
        {hwoz}
\bibfield{author}{\bibinfo{person}{Sruthi Viswanathan},
  \bibinfo{person}{Behrooz Omidvar-Tehrani}, \bibinfo{person}{Adrien Bruyat},
  \bibinfo{person}{Fr\'{e}d\'{e}ric Roulland}, {and}
  \bibinfo{person}{Antonietta~Maria Grasso}.} \bibinfo{year}{2020}\natexlab{b}.
\newblock \showarticletitle{Hybrid Wizard of Oz: Concept Testing a Recommender
  System}. In \bibinfo{booktitle}{\emph{Extended Abstracts of the 2020 CHI
  Conference on Human Factors in Computing Systems}} (Honolulu, HI, USA)
  \emph{(\bibinfo{series}{CHI EA '20})}. \bibinfo{publisher}{Association for
  Computing Machinery}, \bibinfo{address}{New York, NY, USA},
  \bibinfo{pages}{1–7}.
\newblock
\showISBNx{9781450368193}
\urldef\tempurl%
\url{https://doi.org/10.1145/3334480.3383097}
\showDOI{\tempurl}


\end{thebibliography}

\end{document}